# Physics departments should discuss sexual harassment
## —But consider this first

Linda E. Strubbe[1], Electra Eleftheriadou[2], Sarah B. McKagan[3],
Adrian M. Madsen[3], Dimitri R. Dounas-Frazer[4]

1. Kansas State University; 2. University of British Columbia;
3. American Association of Physics Teachers; 4. Western Washington University

In April 2019, Aycock et al. published "Sexual harassment reported by undergraduate female physicists" in Phys. Rev. PER. Libarkin wrote an accompanying editorial for APS called "Yes, Sexual Harassment Still Drives Women Out of Physics." The main finding presented is that 3/4 of undergraduate women in physics in the U.S. report experiencing sexual harassment. Gender minorities[1] experience high rates of harassment also (NASEM Report, APS LGBT Climate Report).

Many physics departments and research groups will want to discuss these articles and findings, with the ultimate goal of making our field a positive and harassment-free place for women and gender minorities. Of course, talking about these findings is crucial towards that goal. However, we want to point out that there are major challenges inherent in having conversations about these articles that departments should address before having these conversations.

As departments and research groups enter into these discussions, they need to recognize that most of the women and gender minorities in the department have experienced and may currently be experiencing sexual harassment and bullying, possibly even perpetrated by other participants in those conversations. For women and gender minorities (and potentially members of other marginalized groups), these conversations are likely to be triggering and require significant emotional labor, and may likely cause

---

[1] A gender minority is an individual whose gender is not as widely represented as others in society; for example, a trans woman is a gender minority. There has been a recent trend to use the acronym GSM, short for Gender and Sexual Minorities, as an inclusive way to refer to persons who self-identify as lesbian, gay, bisexual, transgender, queer, questioning, intersex, as well as other sexual and gender minorities. This terminology obviates the need to list each sub-category which could risk omitting some sub-categories in the process (APS LGBT Climate Report).

Although the focus of sexual harassment discussions is often on the experiences of straight cis-women, GSM physicists also experience high rates of sexual harassment. Sexual harassment is often experienced differently by gender minorities, sexual minorities, and straight cis-women. In the APS LGBT Climate Report, gender minorities reported the most adverse climate in physics, relative to sexual minorities. The report also found that GSM women experienced exclusionary behavior at three times the rate of GSM men. For these reasons, we keep focus in the main text on sexual harassment experienced by women and gender minorities, with acknowledgement that sexual minorities face high levels of harassment as well.

We add that rates of harassment are also particularly high among women of color, for whom harassment may be sexual and racial in nature. In a recent study across racial and gender categories, Clancy et al. (2017) found that women of color in astronomy and planetary science experienced the highest rates of negative workplace experiences. Understanding intersectionality (i.e., the experiences of individuals with two or more marginalized identities) is a crucial part of addressing harassment in our field.

them a level of harm. Following a parallel with anti-racism dialogues (Leonardo & Porter 2010), these discussions may even be unavoidably unsafe for women and gender minorities. For these reasons, discussing Aycock et al.'s findings the way one might discuss a regular paper for journal club or agenda item for a faculty meeting will not work.

Here are a few questions we would encourage organizers of discussions about these papers to reflect on and do reading about.

- What harms might women and gender minorities experience as a consequence of participating in the discussion? What will you do to reduce these potential harms? How will you make it safe for women and gender minorities to participate?
- What is the intended outcome of the discussion? Can you envision an outcome of women and gender minorities participating that would be significant enough that they might choose to open themselves to this potential for harm?
- What role is appropriate for women and gender minorities to play in a discussion? Does everyone explicitly agree to that role? Are there conversations that men should have in men-only environments?
- Is it appropriate to explicitly invite or ask women and gender minorities to participate? What possible power dynamics come into play when people are asked to participate?
- How might intersectional individuals (i.e., those with two or more marginalized identities) experience the discussion differently from white cis-women? How will the discussion include intersectional experiences of harassment?

Two articles that explain some of these challenges further are:

- [On Terms: It's Not My Job to Educate You](#), by Cassandra Lorimer
- [21 Reasons Why It Is Not My Responsibility As a Marginalized Individual to Educate You About My Experience](#), by Elan Morgan

We encourage anyone who wants to discuss these issues to start by reading about harassment, oppression, and difficult conversations. There are many resources accessible online and in libraries. Experts in anti-sexism work and other contexts (e.g., anti-racism work) have thought carefully about how to have difficult, potentially triggering conversations, and when it is and is not appropriate for members of dominant and marginalized groups to have these conversations together. These ideas should be brought to bear explicitly when the physics community discusses these important articles.

Here are a few ideas to help departments start thinking about how to facilitate discussions of sexual harassment.

- Educate yourself about harassment, oppression, and difficult conversations. Find readings on your own, follow suggestions from colleagues, attend talks and workshops on campus and in your community. Reflect on and think critically about what you read and learn.
- Seek out and partner with staff at centers on your campus, such as Women's Resource Centers, Ethnic Student Centers, Gender and Sexuality Centers, or Equity & Inclusion Offices. Staff in

- those centers are experts at facilitating conversations about sexual harassment of women and gender minorities, including those who are Indigenous or non-Indigenous people of color.
- Reach out to other departments on your campus such as Departments of Gender Studies, Queer Studies, and Ethnic Studies. Co-think with faculty in those departments about how to facilitate conversations about sexual harassment in physics.
- Consider hiring an external facilitator to guide a sequence of discussions over the course of multiple semesters. Sexual harassment is a cultural issue, and cultural change is a slow process. External facilitators could support departmental dialogues and possibly also the establishment of new anti-sexist policies or traditions.
- Think about how you will enable access to mental health professionals during and after a discussion. This could include reminding participants of student or employee counseling services they may be entitled to.
- In advance of a discussion, make sure participants have a good understanding of what they should expect in the discussion, so they can mentally prepare. At the beginning of the discussion, explicitly set ground rules for conversation.
- Consider the discussion space carefully. This includes ensuring there are facilities available for participants if they need to take time out from the main discussion space. For example, you could arrange two private auxiliary rooms: a quiet room with comfortable seating and snacks, and a room with a counselor for participants who need someone to talk to right away.
- Consider alternative formats for discussions, such as having groups where men can discuss the article and what they will do about it without asking or expecting women and gender minorities to be present. This is parallel to the idea of [caucus groups](#) in anti-racism work.
- Familiarize yourself with your institution's policy and procedures around sexual harassment, so you can share that information as part of a discussion.

We close by acknowledging that this is indeed difficult. Departments need to think very carefully about how to make safe and productive the conversations our field critically needs to have about sexual harassment and the results in Aycock et al., Clancy et al., the NASEM report and APS LGBT Climate report. But being difficult does not mean our community should give up. Here we have offered a few considerations and ideas for how to make these discussions productive, and our community and institutions have many more ideas and resources. These can be brought to bear in departmental discussions, and in the greater work our field needs to engage in to create a harassment-free environment for all.

**Acknowledgements:**
We are grateful to Eleanor Sayre and Lucy Buchanan-Parker for helpful suggestions on this letter. We would like to acknowledge that much of the writing of this letter took place in Vancouver, British Columbia, on the unceded traditional territory of the xwməθkwəy̓əm (Musqueam), Skwxwú7mesh (Squamish), and Səl̓ílwətaʔ/Selilwitulh (Tsleil-Waututh) First Nations.